\documentclass[prd,aps,showpacs,nofootinbib,eqsecnum]{revtex4}
\usepackage{amssymb}
\usepackage{amsmath}

\def\Lap{{\mathop{\Delta}\limits^{(3)}}}

\newcommand{\bear}{\begin{array}}  \newcommand{\eear}{\end{array}}
\newcommand{\bea}{\begin{eqnarray}}  \newcommand{\eea}{\end{eqnarray}}
\newcommand{\beq}{\begin{equation}}  \newcommand{\eeq}{\end{equation}}
\newcommand{\bef}{\begin{figure}}  \newcommand{\eef}{\end{figure}}
\newcommand{\bec}{\begin{center}}  \newcommand{\eec}{\end{center}}

\newcommand{\bib}{\bibitem}

\newcommand{\Eqn}[1]{&\hspace{-0.2em}#1\hspace{-0.2em}&}

\def\APJ#1#2#3{Astrophys. J. {\bf #1}, #2 (19#3)}
\def\APJJ#1#2#3{Astrophys. J. {\bf #1}, #2 (20#3)}

\def\ARAA#1#2#3{Annu. Rev. Astron. Astrophys. {\bf#1}, #2 (19#3)}
\def\ARAAA#1#2#3{Annu. Rev. Astron. Astrophys. {\bf#1}, #2 (20#3)}

\def\IJMPDD#1#2#3{Int. J. Mod. Phys. D {\bf #1}, #2 (20#3)}

\def\MNRASS#1#2#3{Mon. Not. R. Astron. Soc. {\bf #1}, #2 (20#3)}
\def\MPLA#1#2#3{Mod. Phys. Lett. A {\bf #1}, #2 (19#3)}

\def\PLB#1#2#3{Phys. Lett. B {\bf #1}, #2 (19#3)}
\def\PLBB#1#2#3{Phys. Lett. B {\bf #1}, #2 (20#3)}

\def\PRD#1#2#3{Phys. Rev. D {\bf #1}, #2 (19#3)}
\def\PRDD#1#2#3{Phys. Rev. D {\bf #1}, #2 (20#3)}

\def\PRL#1#2#3{Phys. Rev. Lett. {\bf#1}, #2 (19#3)}
\def\PRLL#1#2#3{Phys. Rev. Lett. {\bf#1}, #2 (20#3)}

\def\PRTT#1#2#3{Phys. Rep. {\bf#1}, #2 (20#3)}

\def\PTPP#1#2#3{Prog. Theor. Phys. {\bf #1}, #2 (20#3)}

\def\RPP#1#2#3{Rep. Prog. Phys. {\bf #1}, #2 (19#3)}

\def\RMPP#1#2#3{Rev. Mod. Phys. {\bf #1}, #2 (20#3)}

\def\Vec#1{\mbox{\boldmath $#1$}}
\def\Vecs#1{\mbox{\boldmath\tiny $#1$}}

\begin{document}

\title{Large-scale magnetic fields in the inflationary universe
      }

\author{Kazuharu Bamba\footnote{
Research Fellow of the Japan Society for the Promotion of Science}
\footnote{E-mail: bamba@yukawa.kyoto-u.ac.jp} 
and Misao Sasaki\footnote{E-mail: misao@yukawa.kyoto-u.ac.jp}}
\affiliation{
Yukawa Institute for Theoretical Physics, 
Kyoto University, 
Kyoto 606-8502, 
Japan}

%\date{\today}

%%%%%%%%%%%%%%%%%%%%%
\begin{abstract}
The generation of 
large-scale magnetic fields is studied in inflationary cosmology.  
We consider the violation of the conformal invariance of the Maxwell
field by dilatonic as well as non-minimal gravitational
couplings. 
We derive a general formula for the spectrum of
large-scale magnetic fields for a general form of the coupling term
and the formula for the spectral index.
The result tells us clearly the (necessary) condition
for the generation of magnetic fields with sufficiently large amplitude.

\end{abstract}
%%%%%%%%%%%%%%%%%%%%%

%----------------------------
\hspace{13.0cm} YITP-06-61
%----------------------------

\maketitle

%%%%%%%%%%%%%%%%%%%%%%%%%%%
%%%  Sec. I 
%%%%%%%%%%%%%%%%%%%%%%%%%%%
\section{Introduction}

It is well known that there exist magnetic fields with the field strength 
$\sim 10^{-6}$G on $1-10$kpc scale in galaxies of all types 
(for detailed reviews of cosmic magnetic fields, 
see 
\cite{Sofue, Kronberg1, Grasso, Carilli1, Widrow, Giovannini1, Semikoz1}) 
and in galaxies at cosmological distances \cite{Kronberg2}. 
Furthermore, 
magnetic fields in clusters of galaxies with the field strength 
$10^{-7}-10^{-6}$G on 10kpc$-$1Mpc scale 
have been observed \cite{Kim1}.   
It is very interesting that magnetic fields in clusters of 
galaxies are as strong as galactic ones and that the coherence scale may be 
as large as $\sim$Mpc.  The origin of these magnetic fields is not well 
understood yet.  
Galactic dynamo mechanisms \cite{EParker} have been proposed 
to amplify very weak seed magnetic fields up to $\sim 10^{-6}$G.  
In fact, however, galactic dynamo mechanisms require initial 
seed magnetic fields to feed on, 
and the effectiveness of the dynamo amplification mechanism 
in galaxies at high redshifts or 
clusters of galaxies is not well established.  

Proposed scenarios for the origin of cosmic magnetic fields fall 
into two broad categories.  One is astrophysical processes, 
%e.g., plasma instabilities, such as the Weibel instability \cite{PI}, 
e.g., the Biermann battery mechanism \cite{Biermann1} 
and the Weibel instability \cite{PI}, which is a kind of 
plasma instabilities, and 
the other is cosmological processes in the early Universe, 
e.g., the first-order cosmological electroweak phase 
transition (EWPT) \cite{Baym}, quark-hadron phase transition (QCDPT) 
\cite{Quashnock} (see also \cite{Boyanovsky1}), 
and the generation of the magnetic fields from primordial 
density perturbations before the epoch of recombination 
\cite{Ichiki1, Siegel1, Gopal1, Berezhiani1}.  
It is difficult, however, for these processes to generate 
the magnetic fields on megaparsec scales with the sufficient field strength 
to account for the observed magnetic fields in galaxies and clusters of 
galaxies without requiring any dynamo amplification.  

The problem is essentially the same as the problem of the large-scale
structure formation for which one needed the adiabatic
curvature perturbation of non-negligible amplitude on superhorizon scales,
the so-called horizon problem.
Then, just as this problem is solved by the inflationary cosmology, it is 
natural to look for the possibility of generating 
such a large-scale magnetic field during inflation~\cite{Turner}.
However, the Friedmann-Robertson-Walker (FRW) universe
is conformal flat and the Maxwell theory is conformal invariant. 
Therefore, the conformal invariance must be broken to 
generate non-trivial magnetic fields.\footnote{In Ref.~\cite{Maroto1},
the effect of the breaking of the conformal flatness
due to scalar metric perturbations at 
the end of inflation is discussed.} 
Various conformal symmetry breaking mechanisms have been proposed so
far~\cite{Turner, RF^2, Ratra, Scalar, Charged-Scalar, 
ScalarED, Amplification, 
Dolgov1, Bertolami1, Gasperini1, Prokopec1, Enqvist1, 
Bertolami2, Ashoorioon1}. 

Bamba and Yokoyama studied the case of the dilaton electromagnetism 
in Ref.~\cite{Bamba1}, where the Lagrangian is in the
form $f(\Phi)F^2$.
The existence of the dilaton is motivated by string theory, and 
such a coupling is reasonable in the light of higher-dimensional 
theories. The case of the coupling function $f(\Phi)=e^{-\lambda\kappa\Phi}$,
where $\kappa=\sqrt{8\pi G}$,
 was first analyzed by Ratra~\cite{Ratra}, 
in which the inflaton and the dilaton are identified.  
Bamba and Yokoyama then showed that magnetic fields with 
the current strength as large as $10^{-10}$G on 1 Mpc scale 
could be generated~\cite{Bamba1}.  
For this to be the case, however, they had to introduce a huge hierarchy 
between the coupling constant of the dilaton to the electromagnetic field 
$\lambda$ and the coupling constant $\tilde\lambda$ in
the dilaton potential, $\lambda/\tilde{\lambda} \gg 1$, 
in order for the spectrum of the generated magnetic fields 
to be sufficiently red so that the amplitude on large scales
can be sufficiently large.

As a possible solution to the above huge 
hierarchy between $\lambda$ and $\tilde\lambda$,  
Bamba and Yokoyama proposed a new scenario~\cite{Bamba2}
by taking account of the effects of 
the stringy spacetime uncertainty relation (SSUR)~\cite{Yoneya}.  
They found that the SSUR of the metric perturbations could lead
to magnetic fields with a nearly scale-invariant spectrum 
even if $\lambda$ and $\tilde\lambda$ are of the same order of magnitude, 
and the amplitude of the generated magnetic fields could be 
as large as $10^{-10}$G on 1Mpc scale at the present time.  

As another mechanism to break the conformal invariance,
non-minimal gravitational couplings were proposed by Turner
and Widrow~\cite{Turner}. Many of the couplings they introduced
break not only the conformal invariance but also the gauge invariance.
The one that looks theoretically more appealing is of the
form $RF^2\,,$
where $R$ is the Ricci scalar. Such a term is
known to arise in curved spacetime due to
 one-loop vacuum-polarization effects~\cite{Drummond}.  

In the present paper, we consider the case 
when there are not only dilatonic and curvature couplings
but also any possible coupling of the form
$I(\Phi,R,\cdots)F^2$, where the dots ($\cdots$) can be
anything.
This paper is organized as follows.  
In Sec.~II, we describe our model and derive the equations of motion.
In Sec.~III, we consider the evolution of the $U(1)$ gauge field 
and that of the magnetic fields in a generic slow-roll inflation model.
We derive formulas for the spectrum of the 
generated magnetic field and the specral index expressed 
very generally in terms of the coupling function $I(\Phi,R,\cdots)$
without specifying its explict form.
These general formulas tell us clearly the necessary condition
for the generation of a sufficiently large amplitude magnetic
field on large scales.
Finally, Sec.~IV is devoted to conclusion.  

%%%%%%%%%%%%%%%%%%%%%%%%%%%
%%%  Sec. II 
%%%%%%%%%%%%%%%%%%%%%%%%%%%
\section{MODEL}

%%%%%%%%%%%%%%%%%%%
%%%  Sec. II A 
%%%%%%%%%%%%%%%%%%%

We consider a conformal violationg Maxwell theory with
the action,
\begin{eqnarray}
S\Eqn{=} 
\int d^{4}x \sqrt{-g} 
\left(-\frac{1}{4}I(\Phi,R,\cdots)\,
F_{\mu\nu}F^{\mu\nu}\right)\,,
\label{eq:conviolation}
\end{eqnarray}
where $I$ is an arbitrary function of
non-trivial background fields at the moment.
Naturally the dots ($\cdots$) in the argument of $I$ 
will include the inflaton if the dilaton $\Phi$ does not 
play the role of the inflaton.

{}From the action~(\ref{eq:conviolation}), 
the electromagnetic field equation is given by
\begin{eqnarray}
-\frac{1}{\sqrt{-g}}{\partial}_{\mu} 
\left[ \sqrt{-g} I F^{\mu\nu} 
\right] \Eqn{=} 0.  
\label{eq:2.6}  
\end{eqnarray}
We assume a spatially flat universe with the metric 
\begin{eqnarray}
{ds}^2 =-{dt}^2 + a^2(t)d{\Vec{x}}^2
= a^2(\eta) ( -{d \eta}^2 + d{\Vec{x}}^2 ), 
\label{eq:2.7}
\end{eqnarray} 
where $a$ is the scale factor and $\eta$ is the conformal time.  
We consider the evolution of the $U(1)$ gauge field in this background.  
Its equation of motion in the Coulomb gauge, 
$A_0(t,\Vec{x}) = 0$ and ${\partial}^jA_j(t,\Vec{x}) =0$, becomes 
\begin{eqnarray}
\ddot{A_i}(t,\Vec{x}) 
+ \left( H + \frac{\dot{I}}{I} 
\right) \dot{A_i}(t,\Vec{x}) 
- \frac{1}{a^2}\Lap\, A_i(t,\Vec{x}) = 0\,,
\label{eq:2.11}
\end{eqnarray}
where $H=\dot a/a$, and a dot denotes
a time derivative, $\dot{~}=\partial/\partial t$.
%%%%%%%%%%%%%%%%%%%
%%%  Sec. III
%%%%%%%%%%%%%%%%%%%
\section{generic slow-roll inflation}
In this section, we consider the evolution of the $U(1)$ gauge field 
and that of the magnetic fields in generic slow-roll inflation.

\subsection{Spectrum from quantum fluctuations}
To begin with, we shall quantize the $U(1)$ gauge field $A_{\mu}(t,\Vec{x})$.  
It follows from the electromagnetic part of our model Lagrangian in 
Eq.~(\ref{eq:conviolation}) that the canonical momenta conjugate to 
$A_{\mu}(t,\Vec{x})$ are given by 
\begin{eqnarray}
{\pi}_0 = 0, \hspace{5mm} {\pi}_{i} = I a(t) \dot{A_i}(t,\Vec{x}).
\label{eq:3.1} 
\end{eqnarray}
We impose the canonical commutation relation 
between $A_i(t,\Vec{x})$ and ${\pi}_{j}(t,\Vec{x})$, 
\begin{eqnarray} 
  \left[ \hspace{0.5mm} A_i(t,\Vec{x}), {\pi}_{j}(t,\Vec{y}) 
  \hspace{0.5mm} \right] = i
 \int \frac{d^3 k}{{(2\pi)}^{3}}
             e^{i \Vecs{k} \cdot \left( \Vecs{x} - \Vecs{y} \right)}
        \left( {\delta}_{ij} - \frac{k_i k_j}{k^2 } \right),
\label{eq:3.2} 
\end{eqnarray}
where $\Vec{k}$ is comoving wave number and $k=|\Vec{k}|$.  
{}From this relation, we obtain the expression for
$A_i(t,\Vec{x})$ as 
\begin{eqnarray} 
  A_i(t,\Vec{x}) = \int \frac{d^3 k}{{(2\pi)}^{3/2}}
  \sum_{\sigma=1,2}\left[ \hspace{0.5mm} \hat{b}(\Vec{k},\sigma) 
        \epsilon_i(\Vec{k},\sigma)A(t,k)e^{i \Vecs{k} \cdot \Vecs{x} }
       + {\hat{b}}^{\dagger}(\Vec{k},\sigma)
       \epsilon_i^*(\Vec{k},\sigma)
         {A^*}(t,k)e^{-i \Vecs{k} \cdot \Vecs{x}} \hspace{0.5mm} \right],
\label{eq:3.3} 
\end{eqnarray}
where $\epsilon_i(\Vec{k},\sigma)$ ($\sigma=1,2$)
are the two orthonormal transverse polarization 
vector, and 
 $\hat{b}(\Vec{k},\sigma)$ and ${\hat{b}}^{\dagger}(\Vec{k},\sigma)$ 
are the annihilation and creation operators which satisfy 
\begin{eqnarray} 
\left[ \hspace{0.5mm} \hat{b}(\Vec{k},\sigma),
 {\hat{b}}^{\dagger}({\Vec{k}}^{\prime},\sigma') \hspace{0.5mm} \right] = 
\delta_{\sigma,\sigma'}
{\delta}^3 (\Vec{k}-{\Vec{k}}^{\prime}), \hspace{5mm}
\left[ \hspace{0.5mm} \hat{b}(\Vec{k},\sigma),
 \hat{b}({\Vec{k}}^{\prime},\sigma')
\hspace{0.5mm} \right] = 
\left[ \hspace{0.5mm} 
{\hat{b}}^{\dagger}(\Vec{k},\sigma),
 {\hat{b}}^{\dagger}({\Vec{k}}^{\prime},\sigma')
\hspace{0.5mm} \right] = 0.
\label{eq:3.4} 
\end{eqnarray}
It follows from Eq.~(\ref{eq:2.11}) that the mode function $A(k,t)$ 
satisfies the equation 
\begin{eqnarray} 
\ddot{A}(k,t) + \left( H + \frac{\dot{I}}{I} \right) 
               \dot{A}(k,t) + \frac{k^2}{a^2} A(k,t) = 0, 
\label{eq:3.5}  
\end{eqnarray} 
with the normalization condition,
\begin{eqnarray} 
A(k,t){\dot{A}}^{*}(k,t) - {\dot{A}}(k,t){A^{*}}(k,t)
= \frac{i}{I a}\,.
\label{eq:3.6} 
\end{eqnarray}
Replacing the independent variable $t$ to $\eta$, we find that 
Eq.~(\ref{eq:3.5}) becomes 
\begin{eqnarray}
A^{\prime \prime}(k,\eta) + 
\frac{I^{\prime}}{I} A^{\prime}(k,\eta) 
+ k^2 {A}(k,\eta) = 0,  
\label{eq:3.7}
\end{eqnarray}
where 
the prime denotes differentiation with respect to the conformal time $\eta$.

Although it is impossible to obtain the exact solution of Eq.~(\ref{eq:3.7})
for the case when $I$ is given by a general function of $\eta$,
we can obtain an approximate solution with sufficient accuracy
by using the WKB approximation on subhorizon scales and the long
wavelength approximation on superhorizon scales, and match these
solutions at the horizon crossing.

In the exact de Sitter background, we have $a=1/(-H\eta)$
where $H$ is the de Sitter Hubble parameter. Hence the
horizon-crossing, which is defined by $H=k/a$, is given by
$-k\eta=1$. The subhorizon (superhorizon) scale corresponds
to the region $k|\eta|\gg1$ ($k|\eta|\ll1$).
We expect this to be also a sufficiently good definition for
the horizon crossing for general slow-roll, almost exponential
inflation.

Then the WKB subhorizon solution is given by
\begin{eqnarray}
A_{\mathrm{in}} (k,\eta) = 
\frac{1}{\sqrt{2k}} I^{-1/2} e^{-ik\eta}, 
\label{eq:subhsol} 
\end{eqnarray} 
where we have assumed that the vacuum in the short-wavelength 
limit is the standard Minkowski vacuum. 

The solution on superhorizon scales can be obtained by
using the longwavelength expansion,
\begin{eqnarray}
A_{\mathrm{out}} = A_0(\eta) + k^2 A_1(\eta) + O(k^4). 
\label{eq:Aout} 
\end{eqnarray} 
Substituting Eq.~(\ref{eq:Aout}) into Eq.~(\ref{eq:3.7}),
we obtain 
\begin{eqnarray} 
A_0^{\prime \prime} + \frac{I^{\prime}}{I} A_0^{\prime} 
\Eqn{=} 0,  
\label{eq:kzero} 
\\
A_1^{\prime \prime} + \frac{I^{\prime}}{I} A_1^{\prime} + 
A_0 \Eqn{=} 0.  
\label{eq:ksquare} 
\end{eqnarray} 
Let the two independent solutions for $A_{\mathrm{out}}$
be $u$ and $v$. For definiteness, we set the boundary conditions to be
$u\to1$ and $v\to0$ as $\eta\to\eta_{\mathrm{R}}$, where  $\eta_{\mathrm{R}}$
is the conformal time at the time of reheating after inflation.
{}From Eq.~(\ref{eq:kzero}), the lowest order solutions 
are given by 
\begin{eqnarray} 
u_0 \Eqn{=} 1,
\label{eq:grow} \\
v_0 \Eqn{=} \int_{\eta}^{{\eta}_{\mathrm{R}}} 
\frac{1}{I \left( \Tilde{\eta} \right)} d \Tilde{\eta}\,.
\label{eq:decay} 
\end{eqnarray}
The solutions accurate to $O(k^2)$ can 
be obtained by the iteration using Eq.~(\ref{eq:ksquare}). 
We find
\begin{eqnarray} 
u \Eqn{=} u_0+k^2u_0\int_{\eta}^{{\eta}_{\mathrm{R}}}d\eta'
I(\eta')\int_{\eta'}^{\eta}
\frac{d\eta''}{I(\eta'')}\,,
\nonumber\\
v \Eqn{=} v_0+k^2\int_{\eta}^{{\eta}_{\mathrm{R}}}d\eta'
v_0(\eta')I(\eta')\int_{\eta'}^{\eta}
\frac{d\eta''}{I(\eta'')}\,.
\end{eqnarray}
With these solutions, the general superhorizon solution is given by
\begin{eqnarray} 
A_{\mathrm{out}}=Cu+Dv, 
\label{eq:Aoutsol} 
\end{eqnarray}
where $C$ and $D$ are constant.

We match the subhorizon solution (\ref{eq:subhsol}) with
the superhorizon solution (\ref{eq:Aoutsol}) at the horizon crossing
$\eta=\eta_k\approx-/k$ to determine the constants $C$ and $D$.
The junction conditions are 
\begin{eqnarray} 
A_{\mathrm{in}} \left( \eta_k\right)
\Eqn{=}
A_{\mathrm{out}} \left( \eta_k\right), 
\label{eq:3.18} \\[3mm]
A_{\mathrm{in}}^{\prime} 
\left( \eta_k\right)
\Eqn{=}
A_{\mathrm{out}}^{\prime} 
\left( \eta_k \right)\,. 
\label{eq:3.19}
\end{eqnarray}
The determination of $C$ and $D$ accurate to $O(k^2)$ 
are given in Appendix. It turns out that $O(k^2)$ corrections
are small unless the function $I$ is non-monotonic and sharply decreases 
on superhorizon scales. 
Hence we neglect the $O(k^2)$ corrections here and approximate
$A_{\mathrm{out}}$ by the lowest order solution,
\begin{eqnarray} 
A_{\mathrm{out}} = 
C(k) + D(k) \int_{\eta}^{{\eta}_{\mathrm{R}}} 
\frac{1}{I \left( \Tilde{\eta} \right)} 
d \Tilde{\eta}. 
\label{eq:superhsol} 
\end{eqnarray} 

Substituting Eqs.\ (\ref{eq:subhsol}) and (\ref{eq:superhsol}) into 
Eqs.\ (\ref{eq:3.18}) and (\ref{eq:3.19}), we obtain 
\begin{eqnarray}  
C(k) \Eqn{=} 
\left.
\frac{1}{\sqrt{2k}} I^{-1/2}  
\left[
1- \left( \frac{1}{2} I^{\prime} + i k I \right) 
\int_{\eta}^{{\eta}_{\mathrm{R}}} 
\frac{1}{I(\eta')} d\eta'\right] e^{-ik\eta} 
\right|_{\eta = \eta_k}, 
\label{eq:C} \\[3mm] 
D(k) \Eqn{=} 
\left.
\frac{1}{\sqrt{2k}} I^{-1/2}  
\left( \frac{1}{2} I^{\prime} + i k I \right) 
e^{-ik\eta} 
\right|_{\eta = \eta_k}.
\label{eq:D} 
\end{eqnarray}  
Neglecting the decaying mode solution, it then 
follows from Eqs.~(\ref{eq:superhsol}) that 
$|A(k,\eta)|^2$ at late times is given by 
\begin{eqnarray}
\left|A(k,\eta)\right|^2&=&|C(k)|^2
=\frac{1}{2kI(\eta_k)}
\left|1- \left(\frac{1}{2}\frac{I^{\prime}(\eta_k)}{kI(\eta_k)}
+ i\right)
e^{-ik\eta_k}k\int_{\eta_k}^{{\eta}_{\mathrm{R}}}
\frac{I(\eta_k)}{I(\eta')}d\eta'\,\right|^2\,.
\label{eq:Aspec}
\end{eqnarray}
The proper magnetic field is given by 
\begin{eqnarray}
{B_i}^{\mathrm{proper}}(t,\Vec{x})
    = a^{-1}B_i(t,\Vec{x}) = a^{-2}{\epsilon}_{ijk}{\partial}_j A_k(t,\Vec{x}),
\label{eq:Bprop}
\end{eqnarray} 
where $B_i(t,\Vec{x})$ is the comoving magnetic field, and 
${\epsilon}_{ijk}$ is the totally antisymmetric tensor
(\hspace{0.5mm}${\epsilon}_{123}=1$\hspace{0.5mm}).  
Thus the spectrum of the magnetic field is given by 
\begin{eqnarray}
|{B}^{\mathrm{proper}}(k,\eta)|^2  
=2\frac{k^2}{a^4}|A(k,\eta)|^2
=2\frac{k^2}{a^4}|C(k)|^2\,,
\label{eq:Bspec}
\end{eqnarray}  
where the factor 2 comes from the two polarization degrees of freedom.
Thus the energy density of the generated magnetic field per
unit logarithmic interval of $k$ is given by
\begin{eqnarray}
\rho_B(k,\eta)\equiv 
\frac{1}{2} 
\frac{4\pi k^3}{(2\pi)^3}|{B}^{\mathrm{proper}}(k,\eta)|^2 I(\eta) 
=\frac{k|C(k)|^2}{2\pi^2}\frac{k^4}{a^4} I(\eta),
\label{eq:Bapprox}
\end{eqnarray}
and the density parameter per unit logarithmic interval of $k$ 
after reheating becomes 
\begin{eqnarray}
\Omega_B(k,\eta)=\frac{\rho_B(k,\eta_{\mathrm{R}})}
{\rho(\eta_{\mathrm{R}})} 
\frac{I(\eta)}{I(\eta_{\mathrm{R}})} 
=\frac{k^4}{T_{\mathrm{R}}^4a_{\mathrm{R}}^4}\frac{15k|C(k)|^2}
{N_{\mathrm{eff}}\pi^4} I(\eta)\,;
\quad 
\rho(\eta_{\mathrm{R}})=N_{\mathrm{eff}}\frac{\pi^2}{30}T_{\mathrm{R}}^4\,,
\label{OmegaB}
\end{eqnarray}
where $T_{\mathrm{R}}$ is the reheating temperature, 
$a_{\mathrm{R}}$ is the scale factor at reheating, and
$N_{\mathrm{eff}}$ is the effective massless degrees of freedom
(2 for photons) which are thermalized at reheating.
The spectral index of $\Omega_B(k)$ is then given by
\begin{eqnarray}  
n_B\equiv \frac{d\ln\Omega_B(k)}{d\ln k}
=4+\frac{d \ln k|C(k)|^2}{d \ln k}\,.
\label{eq:nB}
\end{eqnarray} 

\subsection{Semi-quantitative estimate}
Using Eq.~(\ref{eq:Aspec}) for $|C(k)|^2$, one may derive a general
expression for the density parameter and the spectral index of the
magnetic field in terms of $I$. However, the result will be very complicated
and hence will not be very illuminating. To gain more insight into
semi-quantitative nature of the generated magnetic field, let us
therefore a specific form for the function $I$.
We set
\begin{eqnarray}
I(\eta)=I_*\left(\frac{\eta}{\eta_*}\right)^{-\alpha}\,,
\label{eq:alpha}
\end{eqnarray}
where $\eta_*$ is some fiducial time during inflation, and $\alpha$ is
a constant. As we shall see shortly, in order to obtain a cosmologically
interesting result, we need a monotonically increasing $I$ during the
stage of inflation when the relevant comoving scales leave the horizon. 
So, we assume $\alpha>0$ in the following.
Then $|C|^2$ in Eq.~(\ref{eq:Aspec}) can be explicitly evaluated to be
\begin{eqnarray}
k|C|^2=\frac{1}{2I(\eta_k)}\left|1-\frac{\alpha+2i}{2(\alpha+1)}
e^{-ik\eta_k}\right|^2 \equiv\frac{{\cal C}}{2I(\eta_k)}\,,
\label{powerC}
\end{eqnarray}
where $k\eta_k=O(1)$ and ${\cal C}$ is a constant of order unity.

Then, inserting the above equation to Eqs.~(\ref{OmegaB})
and (\ref{eq:nB}), we obtain the explicit expressions for the
density parameter and the spectral index at present $\eta_0$: 
\begin{eqnarray}
\Omega_B(k,\eta_0)
=\frac{k^4}{T_{\mathrm{R}}^4a_{\mathrm{R}}^4}\frac{15{\cal C}}
{2N_{\mathrm{eff}}\pi^4I_*}
\left(\frac{\eta_k}{\eta_*}\right)^\alpha\,,
\qquad
n_B=4-\alpha\,,
\label{Bkmodel}
\end{eqnarray}
where $I(\eta_k)\propto k^\alpha$ and $I(\eta_0) = 1$ 
have been used.
We see that for a spectrum with a sufficiently large amplitude,
$I_*$ must be sufficiently small, and for a spectral index
close to a scale-invariant one, we need $\alpha\sim4$.

To proceed further to make a more quantitative estimate,
let us assume that the reheating takes place almost instantaneously
so that $\eta_R$ can be identified with the conformal time at the
end of inflation, and $I(\eta)$ is given by Eq.~(\ref{eq:alpha})
until the end of inflation.
Noting that 
$a_{\mathrm{R}}^2\eta_{\mathrm{R}}^2 \approx H_{\mathrm{R}}^{-2}$ 
and $3H_{\mathrm{R}}^2=\rho_{\mathrm{R}}/M_{pl}^2$,
where $M_{pl}=1/\sqrt{8\pi G}$, from Eq.~(\ref{OmegaB}) we obtain
\begin{eqnarray}
\Omega_B(k,\eta_0)
={\cal C}\,\frac{N_{eff}}{1080}
\left(\frac{T_{\mathrm{R}}}{M_{pl}}\right)^4(-k\eta_{\mathrm{R}})^{4-\alpha}
\frac{1}{I(\eta_{\mathrm{R}})}\,.
\end{eqnarray}
We see that higher the reheating temperature is,
larger the density parameter of the magnetic field becomes.
But given the fact that $T_R$ cannot be made arbitrarily large,
there is a limit to this mechanism.
Instead the most important point of the above formula
is the presence of the factor $1/I(\eta_{\mathrm{R}})$. 
It means one can make $\Omega_B$ arbitrarily large
by choosing $I(\eta_{\mathrm{R}})$ arbitrarily small.
An explicit example of this case was discussed in \cite{Bamba1},
where $I$ was a function of a dilaton which is assumed to be rolling
even after the end of inflation. It was found that
the present magnetic field with rms amplitude as large as
$10^{-9}$\,Gauss can be obtained.

Of course, one should be aware of the fact that it is extremely
difficult to construct a realistic model that gives both
a rapidly increasing coupling function $I$ and 
a very small value of $I$ at reheating at the same time.
Even just one of these two conditions is very hard to realize.

Before closing this section, we mention one important fact.
{}From the above result, one might have an impression that there is an extra
enhancement of the magnetic field from the post-inflationary stage.
If this were true, it would mean the existence of a new enhancement
mechanism. But it is not so.
The reality is that the whole enhancement factor is entirely determined
by the amplitude of $I$ at horizon-crossing during inflation. 
This can be seen from the form of $\Omega_B$ given in Eqs.~(\ref{Bkmodel})
in its original form,
\begin{eqnarray}
\Omega_B(k,\eta_0)
=\frac{k^4}{T_{\mathrm{R}}^4a_{\mathrm{R}}^4}\frac{15{\cal C}}
{2N_{\mathrm{eff}}\pi^4}
\frac{1}{I(\eta_k)}\,.
\end{eqnarray}
The above expression clearly shows that it is indeed the value of $I$ at 
horizon-crossing that determines the final amplitude as well as
the spectral index.

%%%%%%%%%%%%%%%%%%%
%%%  Sec. IV
%%%%%%%%%%%%%%%%%%%
\section{Conclusion}

In the present paper we have studied the generation of large-scale magnetic 
fields in inflationary cosmology, breaking the conformal invariance of 
the electromagnetic fields by introducing 
a coupling $IF^2$ with $I$ depending on non-trivial background fields
that vary in time.
We have derived a general formula for the spectrum of 
large-scale magnetic fields for a general form of the coupling term 
and the formula for the spectral index.
The result shows the necessary condition
for the generation of magnetic fields with large amplitude.
Namely, the coupling function $I$ must be extremely small
and rapidly increasing in time during inflation.
Such a coupling seems very unnatural at first sight. 
Nevertheless, considering the recently much discussed issue of the 
string landscape, which tells us that infinitely many different
low energy theories are possible~\cite{Susskind:2003kw}, 
the naturalness may not be crucial, and it may worth searching
for a model that realizes a successful inflationary universe
and satisfies the necessary condition for the generation of
a large magnetic field.

%%%%%%%%%%%%%%%%%%%%

%%%%%%%%%%%%%%%%%%%%%%%%
%%%  Acknowledgements
%%%%%%%%%%%%%%%%%%%%%%%%
\section*{Acknowledgements}

This work was supported in part by 
the Monbu-Kagakusho 21st century COE Program 
``Center for Diversity and Universality in Physics",
and by JSPS Grants-in-Aid for Scientific Research 
(S) No.~14102004 and (B) No.~17340075.
The work of K.B.\ was also supported by a Monbu-Kagakusho
Grant-in-Aid for JSPS Fellows.

%%%%%%%%%%%%%%%%%%%%%%%%%%%%
% Appendix 
%%%%%%%%%%%%%%%%%%%%%%%%%%%%

\appendix*
%\section{WKB solution of order $k^2$ }
\section{}
%In this appendix, we caluculate the WKB solution of order $k^2$.  
%We determine the constants $C$ and $D$ in Eq.~(\ref{eq:45})
%by using the junction conditions in Eqs.\ (\ref{eq:46}) and (\ref{eq:47}).  

In this appendix, we determine the constants $C$ and $D$ 
in Eq.~(\ref{eq:Aoutsol}) 
by taking into account the terms up to order $k^2$.   
Substituting Eq.~(\ref{eq:Aoutsol}) into 
Eqs.\ (\ref{eq:3.18}) and (\ref{eq:3.19}), we find 
\begin{eqnarray}
\left.
\left(
\begin{array}{c}
A_{\mathrm{in}}   \\ 
A_{\mathrm{in}}^{\prime}   
\end{array}
\right)
\right|_{\eta = \eta_k}
=
\mathcal{M}
\left(
\begin{array}{c}
C \\  
D
\end{array}
\right),  
\label{eq:A1} 
%\\[5mm]
\end{eqnarray}
where
\begin{eqnarray}
\mathcal{M} \equiv 
\left.
\left(
\begin{array}{cc}
1- k^2 I_1          &  I_2 \left( 1 - k^2 I_1 \right) \\
k^2 I_3 / I   &  \left[ -1 + k^2 \left( I_1 + I_2 I_3 \right) \right] / I
\end{array}
\right)
\right|_{\eta = \eta_k}.
\label{eq:A2}
\end{eqnarray}
Here, $I_1 (\eta)$, $I_2 (\eta)$, and $I_3 (\eta)$ are defined by 
\begin{eqnarray} 
I_1 (\eta) \Eqn{\equiv} 
\int_{\eta}^{{\eta}_{\mathrm{R}}} 
\left[
\frac{
\int_{\tilde{\eta}}^{{\eta}_{\mathrm{R}}} 
I \left( \Tilde{\Tilde{\eta}} \right) d \Tilde{\Tilde{\eta}} 
}
{I \left( \tilde{\eta} \right)}
\right] d \tilde{\eta}, 
\label{eq:A3}  \\[3mm]
I_2 (\eta) \Eqn{\equiv}
\int_{\eta}^{{\eta}_{\mathrm{R}}} 
\frac{1}{I \left( \tilde{\eta} \right)} d \tilde{\eta},
\label{eq:A4}  \\[3mm]
I_3 (\eta) \Eqn{\equiv} 
\int_{\eta}^{{\eta}_{\mathrm{R}}} 
I \left( \tilde{\eta} \right) d \tilde{\eta}.  
\label{eq:A5}
\end{eqnarray}

From Eqs.\ (\ref{eq:A1}) and (\ref{eq:A2}), we obtain 
\begin{eqnarray}
\left(
\begin{array}{c}
C \\  
D
\end{array}
\right)
=
\mathcal{M}^{-1}
\left.
\left(
\begin{array}{c}
A_{\mathrm{in}}   \\ 
A_{\mathrm{in}}^{\prime}   
\end{array}
\right)
\right|_{\eta = \eta_k}, 
\label{eq:A6} 
%\\[5mm]
\end{eqnarray}
where
\begin{eqnarray}
\mathcal{M}^{-1} = 
\left.
\frac{1}{\left( 1-k^2 I_1 \right)^2}
\left(
\begin{array}{cc}
1 - k^2 \left( I_1 + I_2 I_3 \right)  &  I I_2 \left( 1-k^2 I_1 \right) \\
k^2 I_3                               &  -I \left( 1- k^2 I_1  \right) 
\end{array}
\right)
\right|_{\eta = \eta_k}.
\label{eq:A7}
\end{eqnarray}

%%%%%%%%%%%%%%%%%%%%%%%%%%%%%%%%%
%% thebibliography environment 
%%%%%%%%%%%%%%%%%%%%%%%%%%%%%%%%%


\begin{thebibliography}{99}

\bib{Sofue}
Y.~Sofue, M.~Fujimoto, and R.~Wielebinski, 
\ARAA{24}{459}{86}.

\bib{Kronberg1}
P.~P.~Kronberg, 
\RPP{57}{325}{94}.

\bib{Grasso}
D.~Grasso and H.~R.~Rubinstein, 
\PRTT{348}{163}{01}. 

\bib{Carilli1}
C.~L.~Carilli and G.~B.~Taylor, 
\ARAAA{40}{319}{02}.

\bib{Widrow}
L.~M.~Widrow, 
\RMPP{74}{775}{02}.

\bib{Giovannini1}
M.~Giovannini, 
\IJMPDD{13}{391}{04}.
                        
\bib{Semikoz1}
V.~B.~Semikoz and D.~D.~Sokoloff, 
\IJMPDD{14}{1839}{05}.

\bib{Kronberg2}
P.~P.~Kronberg, J.~J.~Perry, and E.~L.~H.~Zukowski, 
\APJ{355}{L31}{90};\ {\bf 387}, 528 (1992).

\bib{Kim1}
K.~-T.~Kim, P.~P.~Kronberg, P.~E.~Dewdney, and T.~L.~Landecker, 
\APJ{355}{29}{90};\ 
K.~-T.~Kim, P.~C.~Tribble, and P.~P.~Kronberg, 
\textit{ibid}. {\bf 379}, 80 (1991);\ 
T.~E.~Clarke, P.~P.~Kronberg, and H.~B\"ohringer, 
\textit{ibid}. {\bf 547}, L111 (2001).

\bib{EParker}
E.~N.~Parker, 
\APJ{163}{255}{71};\ 
\textit{Cosmical Magnetic Fields} 
(Clarendon, Oxford, England, 1979);\ 
Ya.~B.~Zel'dovich, A.~A.~Ruzmaikin, and D.~D.~Sokoloff, 
\textit{Magnetic Fields in Astrophysics} 
(Gordon and Breach, New York, 1983).

\bib{Biermann1}
L.~Biermann, 
Z.~Naturforsch.\ {\bf 5a}, 65 (1950);\ 
%L.~Biermann and A.~Schl\"{u}ter, 
%\PR{82}{863}{51};\ 
R.~M.~Kulsrud, R.~Cen, J.~P.~Ostriker, and D.~Ryu, 
\APJ{480}{481}{97};\ 
N.~Y.~Gnedin, A.~Ferrara, and E.~G.~Zweibel, 
\textit{ibid}. {\bf 539}, 505, (2000);\ 
G.~Davies and L.~M.~Widrow, 
\textit{ibid}. {\bf 540}, 755, (2000);\ 
H.~Hanayama \textit{et al}., 
\textit{ibid}. {\bf 633}, 941, (2005).  

\bib{PI}
E.~S.~Weibel,
\PRL{2}{83}{59};\ 
N.~Okabe and M.~Hattori, 
\APJJ{599}{964}{03};\ 
T.~N.~Kato, 
Phys.~Plasmas {\bf 12}, 080705 (2005);\  
Y.~Fujita and T.~N.~Kato, 
\MNRASS{364}{247}{05}. 

\bib{Baym}
G.~Baym, D.~B\"odeker, and L.~McLerran, 
\PRD{53}{662}{96}.

\bib{Quashnock}
J.~M.~Quashnock, A.~Loeb, and D.~N.~Spergel, 
\APJ{344}{L49}{89}.

\bib{Boyanovsky1}
D.~Boyanovsky, H.~J.~de Vega, and M.~Simionato,
\PRDD{67}{123505}{03};\ 
\PRDD{67}{023502}{03}.

\bib{Ichiki1}
K.~Ichiki \textit{et al}.,
%K.~Takahashi, H.~Ohno, H.~Hanayama, and N.~Sugiyama, 
Science {\bf 311}, 827 (2006); 
K.~Takahashi, K.~Ichiki, H.~Ohno, and H.~Hanayama, 
\PRLL{95}{121301}{05}.
%, astro-ph/0603631.

\bib{Gopal1}
R.~Gopal and S.~Sethi, 
\MNRASS{363}{529}{05}.  

\bib{Siegel1}
E.~R.~Siegel and J.~N.~Fry, 
arXiv:astro-ph/0604526. 

\bib{Berezhiani1}
Z.~Berezhiani and A.~D.~Dolgov, 
Astropart.~Phys.\ {\bf 21}, 59 (2004).

%\bib{Durrer}
%R.~Durrer and C.~Caprini,                    
%J.~Cosmol.~Astropart.~Phys.\ {\bf 11}, 010 (2003).
%\bib{Durrer}
%R.\ Durrer and C.\ Caprini,                    
%J.\ Cosmol.\ Astropart.\ Phys.\ 0311, 010 (2003).

\bib{Turner}
M.~S.~Turner and L.~M.~Widrow, 
\PRD{37}{2743}{88}.

%\bib{Linde1}
%A.~D.~Linde, 
%\textit{Particle Physics and Inflationary Cosmology} 
%(Harwood Academic, Chur, Switzerland, 1990);\ 
%K.~A.~Olive, 
%\PRT{190}{307}{90};\ 
%D.~H.~Lyth and A.~Riotto, 
%\textit{ibid}. {\bf 314}, 1, (1999).
%
%\bib{Kolb}
%E.~W.~Kolb and M.~S.~Turner, 
%\textit{The Early Universe} (Addison-Wesley, Redwood City, California, 1990).
%
%%%%%
%\bib{Guth}
%A.~H.~Guth and S.~-Y.~Pi, 
%\PRL{49}{1110}{82};\ 
%S.~W.~Hawking, 
%\PL{115B}{295}{82};\ 
%A.~A.~Starobinsky, 
%\textit{ibid}. {\bf 117B},\ 175\ (1982);\ 
%J.~M.~Bardeen, P.~J.~Steinhardt, and M.~S.~Turner, 
%\PRD{28}{679}{83}.

%\bib{Starobinsky}
%A.~A.~Starobinsky, 
%\JL{30}{682}{79}.

%\bib{Rubakov}
%V.~A.~Rubakov, M.~V.~Sazhin, and A.~V.~Veryaskin, 
%\PL{115B}{189}{82}.
%%%%%

\bib{Maroto1}
A.~L.~Maroto,
\PRDD{64}{083006}{01}.

%\bib{Parker}
%L.~Parker, 
%\PRL{21}{562}{68}.

\bib{RF^2}
F.~D.~Mazzitelli and F.~M.~Spedalieri, 
\PRD{52}{6694}{95};\ 
G.~Lambiase and A.~R.~Prasanna, 
\textit{ibid}.\ {\bf 70}, 063502 (2004). 

\bib{Ratra}
B.~Ratra, 
\APJ{391}{L1}{92};\ 
Report No. GRP-287/CALT-68-1751.

\bib{Scalar}
W.~D.~Garretson, G.~B.~Field, and S.~M.~Carroll, 
\PRD{46}{5346}{92};\ 
D.~Lemoine and M.~Lemoine, 
\textit{ibid}. {\bf 52}, 1955 (1995);\ 
%\PRD{52}{1955}{95};\ 
M.~Gasperini, M.~Giovannini, and G.~Veneziano, 
\PRL{75}{3796}{95};\ 
M.~Giovannini, 
\PRDD{64}{061301}{01};\ 
%\textit{ibid}. {\bf 64}, 061301 (2001);\ 
arXiv:hep-ph/0104214;\ 
arXiv:astro-ph/0212346;\ 
M.~R.~Garousi, M.~Sami, and S.~Tsujikawa, 
\PLBB{606}{1}{05};\ 
M.~A.~Ganjali, 
J.\ High Energy Phys.\ 09 (2005) 004;\ 
M.~M.~Anber and L.~Sorbo, 
J.\ Cosmol.\ Astropart.\ Phys.\ 10 (2006) 018.


\bib{Charged-Scalar}
E.~A.~Calzetta, A.~Kandus, and F.~D.~Mazzitelli, 
\PRD{57}{7139}{98};\ 
A.~Kandus, E.~A.~Calzetta, F.~D.~Mazzitelli, and C.~E.~M.~Wagner,
\PLBB{472}{287}{00};\ 
M.~Giovannini and M.~Shaposhnikov, 
\PRDD{62}{103512}{00}.

\bib{ScalarED}
A.~-C.~Davis, K.~Dimopoulos, T.~Prokopec, and O.~T\"ornkvist, 
\PLBB{501}{165}{01};\ 
K.~Dimopoulos, T.~Prokopec, O.~T\"ornkvist, and A.~C.~Davis, 
\PRDD{65}{063505}{02};\ 
T.~Prokopec, O.~T\"{o}rnkvist, and R.~Woodard, 
\PRLL{89}{101301}{02}.

\bib{Amplification}
B.~A.~Bassett, G.~Pollifrone, S.~Tsujikawa, and F.~Viniegra, 
\PRDD{63}{103515}{01};\ 
F.~Finelli and A.~Gruppuso, 
\PLBB{502}{216}{01}.

\bib{Dolgov1}
A.~D.~Dolgov, 
\PRD{48}{2499}{93}.

\bib{Bertolami1}
O.~Bertolami and D.~F.~Mota, 
\PLB{455}{96}{99}.  

\bib{Gasperini1}
M.~Gasperini, 
\PRDD{63}{047301}{01}.

\bib{Prokopec1}
T.~Prokopec, 
arXiv:astro-ph/0106247.

\bib{Enqvist1}
K.~Enqvist, A.~Jokinen, and A.~Mazumdar, 
J.~Cosmol.~Astropart.~Phys.\ 11 (2004) 001.

\bib{Bertolami2}
O.~Bertolami and R.~Monteiro,
\PRDD{71}{123525}{05}. 

\bib{Ashoorioon1}
A.~Ashoorioon and R.~B.~Mann, 
\PRDD{71}{103509}{05}. 

%%%%%
%\bib{HD}
%M.~Giovannini, 
%\PRDD{62}{123505}{00}; 
%K.~E.~Kunze, 
%\PLBB{623}{1}{05}.
%%%%%

\bib{Bamba1}
K.~Bamba and J.~Yokoyama, 
\PRDD{69}{043507}{04}.

\bib{Bamba2}
K.~Bamba and J.~Yokoyama, 
\PRDD{70}{083508}{04}.

\bib{Yoneya}
T.~Yoneya, 
\MPLA{4}{1587}{89};\ 
M.~Li and T.~Yoneya,
\PRL{78}{1219}{97};\ 
T.~Yoneya, 
\PTPP{103}{1081}{00}.

\bib{Drummond}
I.~T.~Drummond and S.~J.~Hathrell, 
\PRD{22}{343}{80}.  

%%%%%
%\bib{Abbott1}
%L.~F.~Abbott and M.~B.~Wise, 
%\NPB{244}{541}{84}.
%
%\bib{Rubakov}
%V.~A.~Rubakov, M.~V.~Sazhin, and A.~V.~Veryaskin, 
%\PL{115B}{189}{82}.
%
%\bibitem{Spergel06}
%D.~N.~Spergel \textit{et al}., 
%astro-ph/0603449. 
%%%%%
%\cite{Susskind:2003kw}
\bibitem{Susskind:2003kw}
  L.~Susskind,
  %``The anthropic landscape of string theory,''
  arXiv:hep-th/0302219.
  %%CITATION = HEP-TH 0302219;%%
\end{thebibliography}
\end{document}